\documentclass[%
 reprint,
]{revtex4-2}

\usepackage{graphicx}
\usepackage{tikz}

\usepackage[utf8]{inputenc}
\usepackage{siunitx}
\usepackage{bm}
\usepackage[colorlinks]{hyperref} 
\hypersetup{
	linkcolor=blue,
	citecolor=red,
	urlcolor=blue
}
\usepackage{xcolor}
\usepackage[normalem]{ulem}

\usepackage{mathtools}
\usepackage{booktabs} 
\usepackage{multirow}
\usepackage{textcomp} 
\usepackage{amssymb} 
\usepackage[version=4]{mhchem}
\usepackage{wasysym}

\begin{document}


\title{A Systematic Error in the Internal Friction Measurement of Coatings for Gravitational Waves Detectors}

\author{Alex Amato$^1$}\email{a.amato@maastrichtuniversity.nl}
\author{Diana Lumaca$^{2,3}$}
\author{Elisabetta Cesarini$^3$}
\author{Massimo Granata$^4$}
\author{Anaël Lemaître$^5$}
\author{Matteo Lorenzini$^{2,3}$}
\author{Christophe Malhaire$^6$}
\author{Christophe Michel$^4$}
\author{Francesco Piergiovanni$^7$}
\author{Laurent Pinard$^4$}
\author{Nikita Shcheblanov$^5$}
\author{Gianpietro Cagnoli$^1$}
\affiliation{[1] Universit\'e de Lyon, Universit\'e Claude Bernard Lyon 1, CNRS, Institut Lumi\`ere Mati\`ere, F-69622, VILLEURBANNE, France.}
\affiliation{[2] University of Roma Tor Vergata via della ricerca scientifica 1 00133 Roma.}
\affiliation{[3] Istituto Nazionale di Fisica Nucleare, sez. Roma Tor Vergata, via della ricerca scientifica 00133 Roma.}
\affiliation{[4] Laboratoire des Mat\'{e}riaux Avanc\'{e}s - IP2I, CNRS,
Universit\'{e} de Lyon, Universit\'{e} Claude Bernard Lyon 1, F-69622
Villeurbanne, France.}
\affiliation{[5] Navier, Ecole des Ponts, Univ Gustave Eiffel, CNRS, Marne-la-Vallée, France}
\affiliation{[6] Univ Lyon, INSA Lyon, ECL, CNRS, UCBL, CPE Lyon, INL, UMR5270, 69621 Villeurbanne, France}
\affiliation{[7] Università degli Studi di Urbino ‘Carlo Bo’, I-61029 Urbino, Italy.}

\date{\today}

\begin{abstract}
Low internal friction coatings are key components of advanced technologies such as optical atomic clocks and high-finesse optical cavity and often lie at the forefront of the most advanced experiments in Physics. Notably, increasing the sensitivity of gravitational-wave detectors depends in a very large part on developing new coatings, which entails developing more suitable methods and models to investigate their loss angle. In fact, the most sensitive region of the detection band in such detectors is limited by the coating thermal noise, which is related to the loss angle of the coating. Until now, models which describe only ideal physical properties have been adopted, wondering about the use of one or more loss angles to describe the mechanical properties of coatings. Here we show the presence of a systematic error ascribed to inhomogeneity of the sample at its edges in measuring the coating loss angle. We present a model for disk-shaped resonators, largely used in loss angle measurements, and we compare the theory with measurements showing how this systematic error impacts on the accuracy with which the loss model parameters are known.
\end{abstract}

\maketitle

\section{Introduction}
The search for low internal friction coatings lies today at the forefront of various branches of Physics requiring ever more accurate measurements of time and space: optical atomic clocks~\cite{Rosenband1808} and high-finesse optical cavity for quantum devices~\cite{PhysRevLett.120.060601}, laser stabilization~\cite{Kessler2012} and optomechanical resonators~\cite{RevModPhys.86.1391}, or the ground-based gravitational-wave detectors (GWDs)~\cite{accadia2012virgo,0264-9381-32-7-074001}, on which depends the emergence of a new branch of astrophysics. 

GWDs are large interferometers in which massive suspended mirrors play the role of gravitational-field probes~\cite{RevModPhys.86.121}. 
The current GWDs mirror coatings are Bragg reflectors, made of titania-tantala (\ce{TiO2}:\ce{Ta2O5}) and silica (\ce{SiO2}) doublets, two amorphous materials with, respectively, high and low refractive indices, and which are deposited by ion-beam sputtering~\cite{Amato_2018, Amato_2019, Amato_2019_annealing, granata2020amorphous}. After the first successful observation runs~\cite{PhysRevX.9.031040,Collaboration2017}, GWDs are now entering in an upgrade phase to improve their sensitivity. 
One of the most critical tasks is to reduce the coating thermal noise, which limits detection in the central band where the interferometers are the most sensitive~\cite{Harry:06}.

This thermal noise is especially large in coatings due to their strongly out-of-equilibrium nature. But it is not specific to them. 
It is found in bulk glass, where it is believed to arise from thermally driven transitions between metastable atomic configurations resulting from structural disorder.
Thanks to the fluctuation-dissipation theorem, it can be related to internal friction, i.e., the attenuation of mechanical excitations in the bulk of material, a property characterized by loss angles, respectively associated with compressive and shear strains.
Precise and reliable measurements of these properties are hence crucial to characterize thermal noise and provide information about its structural origin~\cite{PhysRevLett.84.2718,0295-5075-80-5-50008,PhysRevMaterials.2.053607,Amato_2020}.

Most experiments used to access the intrinsic elastic moduli and loss angles in coatings rely on the principle of the resonant method~\cite{nowick1972anelastic}, via measurements of the resonance frequencies and internal friction for various types of resonators~\cite{GranataInternalFriction,PhysRevD.93.012007}, the more reliable and most used technique being the so-called Gentle Nodal Suspension (GeNS) for disk-shaped resonators.~\cite{doi:10.1063/1.3124800,0264-9381-27-8-084031,GranataInternalFriction,PhysRevD.93.012007}. In these setups, one may neglect non-viscous dissipation contributions like the thermoelastic loss. 
However, one does not access independently the targeted loss angles, because every resonant mode combines both compressive and shear contributions. These angles can only be estimated after matching experimental resonance data with a loss model, while taking into account the distribution of local strains in every vibration mode. For this purpose, most of the recent GeNS studies of amorphous coatings assume that, non-viscous dissipation being negligible, the total coating dissipation results only from the bulk internal friction of the deposited material, characterized by the two loss angles one seeks to measure~\cite{PhysRevD.87.082001,PhysRevLett.127.071101}.

In this work, we show, on the example of GeNS measurements, that this procedure leads to a severe misestimation of the targeted intrinsic loss angles, due to the existence of excess losses arising from the coating edge. A similar edge effect was previously evidenced in uncoated disks~\cite{cagnoli2017mode}. It should be expected to be the rule in coatings~\cite{TILLI201071} because they are highly prone to the existence of edge inhomogeneity arising, e.g., from spills off at the substrate edge or from tapering on the front side due to the sample holder during deposition. Our work shows that a very slight excess of loss at the edge suffices to strongly affect the measured loss at higher frequencies because modes are then increasingly localized near the disk edge. It results in a large error of the fitted intrinsic loss angles. We also point out that a lack of coating near the disk edge introduces an artefact similar to the existence of a negative edge effect, that must also be taken into account in order to access the desired intrinsic loss angles.

\section{State of the art}
\subsection{The resonant Method}
\label{sec:resonant}

The Gentle Nodal Suspension (GeNS)~\cite{doi:10.1063/1.3124800,0264-9381-27-8-084031,GranataInternalFriction,PhysRevD.93.012007} is one of the most reliable tools to access the mechanical properties of coatings. It proceeds on the principle of the resonant method~\cite{nowick1972anelastic}, via measurements of the resonance frequencies and internal friction of non-coated and coated disks. To interpret measurements of the loss angle made with this device, and access the intrinsic coating loss, it was recently proposed~\cite{cagnoli2017mode} to proceed as follows. First, the system is decomposed into groups of static degrees of freedom, such as shear vs bulk strains, in different subparts of the system, each one being associated with a stored elastic energy and susceptible to various dissipation mechanisms (e.g. structural relaxation and thermoelastic effect). Then, when examining a given resonance of the system, the rate associated with each dissipation mechanism is assumed to be proportional to the averaged stored energy in the corresponding group over a period of vibration for that resonance mode. Thus, a system is divided into several domains (e.g. substrate vs coating, indexed by $i$) and several groups of degrees of freedom (e.g. bulk vs shear strains, indexed by $k$); the average elastic energy of group $(k,i)$ over a period of the considered resonance mode is denoted $E_{i,k}$; and in each group the dissipation rate via the mechanisms $m$ is assumed to be $\phi^k_{i,m}E_{i,k}$. Writing the total energy dissipation rate as the sum of these contributions, the loss angle of the whole (composite) system, associated with the decay of a single mode, reads:
\begin{equation}
\phi_{\text{tot}} = \sum_i\sum_k D_{i,k} \left \{ \sum_m\phi^k_{i,m}\right \}\medspace,
\label{Phi_tot_generic}
\end{equation}
where 
\begin{equation}
    D_{i,k}=\frac{E_{i,k}}{E_{\text{tot}}}\medspace,
\end{equation} 
is the the so-called dilution factor, which depends on the considered resonance mode.

\subsection{The coating loss angle}
Equation~(\ref{Phi_tot_generic}) appears in its simplest form when merely splitting substrate ($s$) from coating ($c$) contributions. Assuming the latter homogeneous and isotropic, and the total energy $E$ entirely distributed between the two part $s$ and $c$, leads to
\begin{align}
\phi_{\text{tot}} &= D_s \phi_s + D_c\phi_c\medspace \notag\\
 &= (1 - D) \phi_s + D\phi_c\medspace, 
\label{eq:semplice}
\end{align}
where $\phi_s$ and $\phi_c$ are the loss angles associated to the internal friction of the substrate and the coating respectively, and $D=D_c=1-D_s$.
This is how the coating loss angle is usually deduced, as
\begin{equation}
\phi_c = \frac{1}{D}\left [ \phi_{\text{tot}}-(1-D)\phi_s \right ] \medspace.
\label{eq:phiCbase}
\end{equation} 
We thus obtain the loss angle of the coating directly from the measurement of the total (coated) sample from which we remove the measurement of the bare substrate, by weighting the contributions with the dilution factor. In the following, we assume that the substrate loss $\phi_s$ does not change during the coating deposition and post-deposition treatments.

Equation~(\ref{eq:phiCbase}) highlights that the measurement of the coating loss angle rests entirely of a precise evaluation of the dilution factor for each resonance mode.

From the geometry of the sample, it is possible to obtain the dilution factor after calculating the energies stored in the either part of the system using finite-element simulations. This procedure, however, requires a detailed knowledge of the mechanical parameters of both substrate and coating, which may be quite limiting when examining new, lesser known, materials. 

Alternatively, the dilution factor may be directly accessed experimentally via the shift of resonance frequencies after coating deposition, as detailed in Appendix~\ref{app:D}. For a thin disk with a homogeneous coating covering the entire surface of the substrate, we have:
\begin{equation}
D = 1- \frac{m_s}{m_{\text{tot}}}\left(\frac{f_s}{f_{\text{tot}}}\right)^2\medspace,
\label{eq:Dilution}
\end{equation} 
where $f_s$, $f_{\text{tot}}$, $m_s$ and $m_{\text{tot}}$ are the resonance frequencies and the mass of the sample before and after coating deposition, respectively. Once $D$ is estimated through equation~(\ref{eq:Dilution}) the coating loss angle $\phi_c$ can be worked out following equation~(\ref{eq:phiCbase}).

\begin{figure*}
	\centering
	\includegraphics[width=0.28\textwidth]{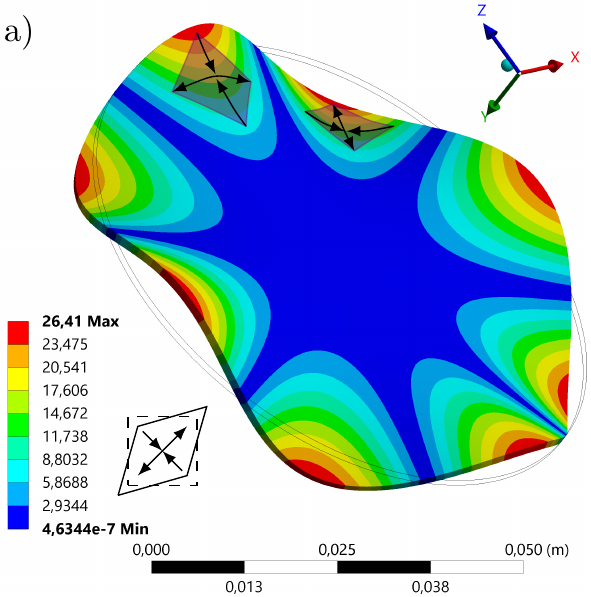}\qquad
	\includegraphics[width=0.3\textwidth]{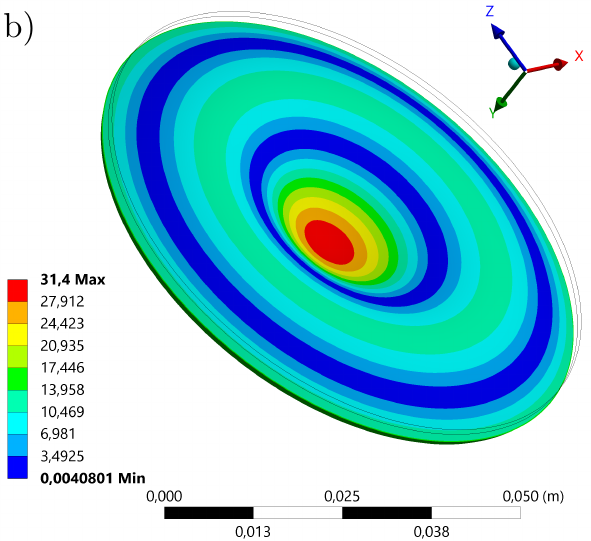}\qquad
	\includegraphics[width=0.3\textwidth]{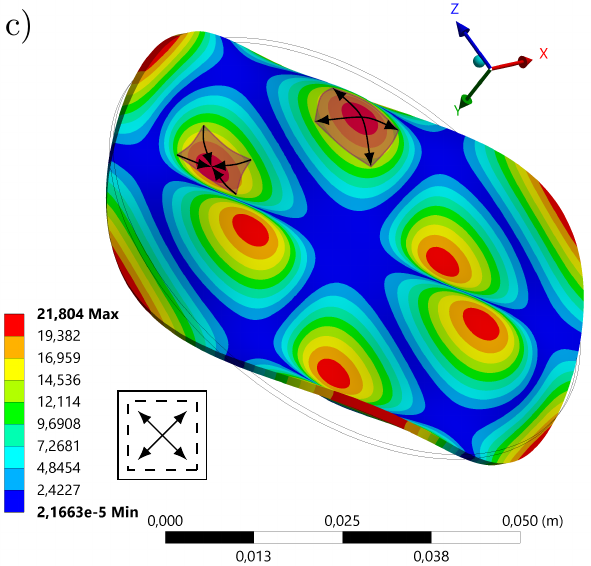}
	\caption{Normal displacement amplitude for a few resonance modes of a 75 mm, 1~mm thick, silica ($\nu=0.16$) disk-shaped resonator, as obtained from finite element simulations. The displacement goes from zero (blue) to the maximum of the deformation during the vibration (red) highlighting the mode shape. A mode $(m,n)$ has $m$ radial nodes and n angular or azimuthal nodes. The three examples show mode (0,4) in panel a), mode (2,0) in panel b) and mode (1,3) in panel c). In modes of family $(0,n)$, like mode (0,4) in panel a), the local curvatures along the radial and orthoradial directions have opposite signs, which causes the associated strains to be primarily deviatoric (shear). In modes having non-zero $m$ values, like the (1,3) reported  in panel c), these two curvatures have the same sign over a significant part of the disk area, which corresponds primarily to compressive (bulk) strains. The sketches at the bottom left of the modes in panels a) and c) represent the dominant local deformation.}
	\label{fig:ansys}
\end{figure*}

\subsection{Bulk and shear}

A first step towards the modeling of coating loss consists in constructing Eq.~(\ref{Phi_tot_generic}) while separating the contributions of different elastic strains. Even though coatings display an evident asymmetry between the normal and longitudinal directions, there are various indications~\cite{Pinard:17,10.1117/12.981766} that, when they grow amorphous, this does not usually induce a significant structural anisotropy~\footnote{There is inevitably a stress asymmetry, since films tend to grow under compressive stresses in the transverse plane, while the free surface introduce a zero stress condition in the normal direction; this, however, does not necessarily introduces a significant structural anisotropy so long as it is equivalent to the effect of a moderate  elastic loading.}. 

This justifies focusing on structurally isotropic coatings, the elastic response of which can be assumed to be fully captured by just two elastic constants. Accordingly, we are then led to expect that the film loss can be fully captured to just two contributions associated with either bulk or shear strains~\cite{ABERNATHY20182282}, in which case Eq.~(\ref{Phi_tot_generic}) reduces to:
\begin{equation}
\phi_c = D_\text{bulk} \phi_\text{bulk}+D_\text{shear} \phi_\text{shear}\medspace,
\label{eq:bulkshear_simple}
\end{equation}
where $D_\text{bulk}$, $D_\text{shear}=1-D_\text{bulk}$ are the bulk and shear dilution factors, which are computed in Appendix~\ref{app:bulkshear}, which can be found in table~\ref{tab:dilution}.

\begin{figure}
	\centering
	\includegraphics[width=0.8\columnwidth]{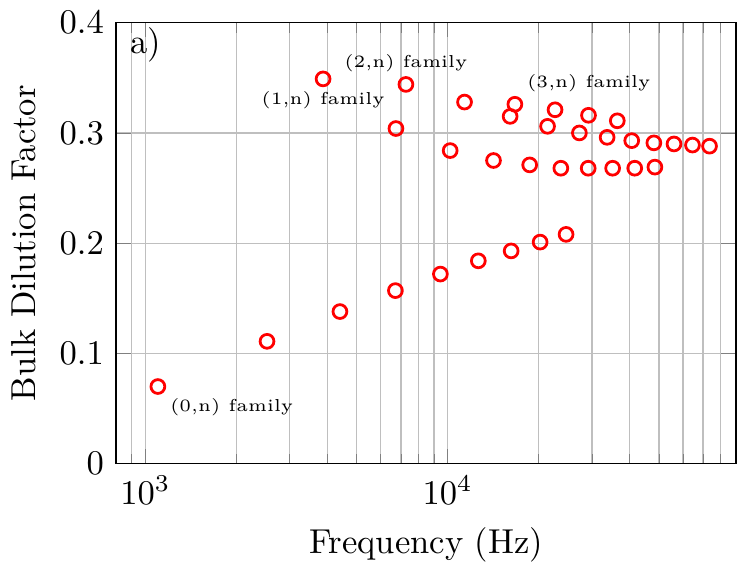}
	\includegraphics[width=0.8\columnwidth]{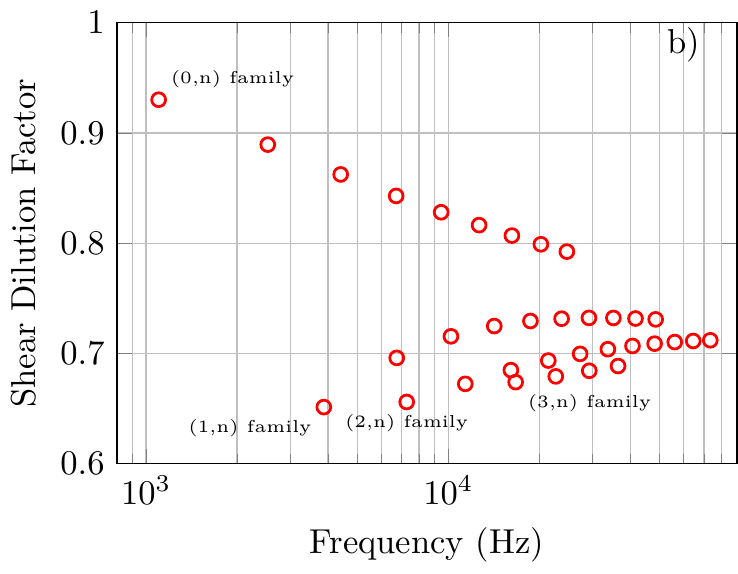}
	\caption{a) Bulk dilution factor $D_\text{bulk}$, b) Shear dilution factor $D_\text{shear}=1-D_\text{bulk}$. The frequency modes are those of a 75 mm, 1~mm thick disk. For these numbers a Poisson's ratio of 0.16 for the mode shape $w(r,\theta)$ and 0.16 for the coating have been considered.}
	\label{fig:Dbulkshear}
\end{figure}

For a disk, the values of $D_\text{bulk}$ and $D_\text{shear}$ only depend on the Poisson ratio, and on the radial and azimuthal numbers characterizing each mode. To illustrate the factors affecting the dilution factors, a few resonance modes of a thin disk are displayed in figure~\ref{fig:ansys}. As shown in panel a), modes having only azimuthal nodes feature mostly shear strains, hence tend to present the largest $D_\text{shear}$ values. In contrast, as illustrated in panel b) and c), as the radial number grows, the modes display increasing areas undergoing compressive strains near antinodes, so that $D_\text{shear}$ tends to decrease with $m$. The resulting dilution factors are displayed in figure~\ref{fig:Dbulkshear} as a function of frequency for the same disk.

This figure illustrates that the values of the dilution factors for different $m$'s lie on different curves, which multiple families of modes resonate over the frequency range of interest. This is why typical coating loss angle data [Eq.~(\ref{eq:bulkshear_simple})] is not a single-valued function of frequency but lie on different curves. In contrast, the intrinsic loss angles $\phi_\text{bulk}$ and $\phi_\text{shear}$ are material properties, hence should be continuous functions of frequency.

\subsection{Loss angle models}

The last step towards the modelling of coating loss consists in considering a proper theory for the dissipation mechanisms. The most common model assumes that dissipation results from two-level systems (or TLS's), i.e. microstructural elements that flip between states separated by a potential energy barrier, thus inducing fluctuations of the local elastic modulus $M$, hence a phase lag between stress and strain. TLS's are typically modelled by asymmetric double-well potentials~\cite{doi:10.1080/01418638108222343}, with distributed parameters, in which case the loss angle reads
\begin{equation}
\phi = \int_0^\infty\int_0^\infty\left(\frac{\delta M}{M}\right)\frac{\omega\tau}{1+(\omega\tau)^2}g(\Delta,V)d\Delta dV\medspace,
\label{eq:TLS}
\end{equation}
where $\delta M$ is the variation of the elastic modulus, $\omega=2\pi f$, $\tau=\tau_0\exp(V/k_BT)/(1+\exp(-\Delta/k_BT))$ the characteristic flip time, and where $g$ is the distribution of the asymmetry $\Delta$ and barrier height $V$ of the asymmetric double-well potential, which accounts for the complexity of the amorphous structure. 

At that stage, the frequency dependence of $\phi$ is entirely encoded in the integration of Eq. \ref{eq:TLS}, using a particular type of $g$. To progress, in common oxide materials like silica and tantala, $\Delta$ and $V$ are typically assumed to be independent and to be, respectively, uniformly and exponentially distributed~\cite{PhysRevB.93.014105,TRAVASSO2009268}. The distribution of $V$ being written:
\begin{equation}
\upsilon(V)=\frac{1}{V_0}e^{-V/V_0}\medspace,
\label{eq:V_Exponential}
\end{equation}
with $V_0$ a material-dependent constant, one then finds the loss angle $\phi$ to scale as a power law of frequency:
\begin{equation}
\phi = \phi_0\left(\omega\tau_0\right)^{\alpha(T)}\medspace,
\label{eq:Phi_TLS}
\end{equation}
with an exponent $\alpha(T)=k_BT/V_0$ that grows linearly with temperature.

Such a power law dependence of $\phi$ has been observed in vitreous silica at low temperatures~\cite{PhysRevB.64.064207,TRAVASSO2009268}. It gives a strong support to the idea that the distribution of $V$ is indeed exponential, yet only in a small energy range corresponding to temperatures up to about 130K. The available evidence coming from this work suggests that at higher temperatures, the exponent $\alpha$ no longer grows with temperature but instead drops. This casts doubts on the relevance of a frequency power law estimate for $\phi$ at room temperature, although such fits are commonly employed to analyze experimental data. Thus, while power law fits of the loss angles remain empirically instructive, for lack of a better option, they should be considered with a degree of skepticism.

Let us also point out that, when power law expressions for the loss angles are used, one sometimes finds Eq.~(\ref{eq:Phi_TLS}) combined with Eq.~(\ref{eq:bulkshear_simple}) as follows:
\begin{equation}
\phi_c = D_\text{bulk} A_1f^{\alpha_1} + D_\text{shear} A_2f^{\alpha_2}\medspace,
\label{eq:bulkshear}
\end{equation}
with distinct coupling constants $A_1$ and $A_2$, and exponents $\alpha_1$ and $\alpha_2$.

Within the TLS model, the constants $A_1$ and $A_2$ result from the material elastic properties, parameters of the TLS distribution and strain couplings; meanwhile, as explained above, the exponents $\alpha_1$ and $\alpha_2$ are expected to be $k_bT/V_0$ with $V_0$ the characteristic energy of the TLS distribution. From this standpoint, the coupling constants $A_1$ and $A_2$ are legitimately distinct, but the two exponents $\alpha_1$ and $\alpha_2$ should be equal. Allowing these two exponents to take different values, as sometimes seen in the literature, amounts to introducing implicitly the assumption that there is not one population of TLS, but two, which couple either to the bulk or to the shear strains. This, we think, is rather dubious as the TLS model should instead be interpreted as representing the existence of a single barrier population that couples to both bulk and shear strains.

\section{Edge effects}
\subsection{Fully coated disk}
\label{sec:Edgedisk}

It was recently shown~\cite{cagnoli2017mode} that, in non-coated silica disk-shaped resonators, the purely azimuthal modes $(0,n)$ present an excess loss that cannot be described by a simple decomposition in bulk and shear strains, but should be attributed to spurious losses near the disk-edge, where purely azimuthal modes display their largest strains. This so-called edge effect is expected to arise from surface inhomogeneities. In coated resonators, these losses contribute to $\phi_s$ and are eliminated when evaluating the coating loss $\phi_c$ via Eq.~(\ref{eq:phiCbase}).

The present paper emphasizes that a similar effect should arise from the coating edge, hence impacting $\phi_c$ measurements. Consider, indeed, the bulk and shear dilution factors of Fig.~\ref{fig:Dbulkshear}: their values for different mode families (i.e. for different values of the radial number $m$) tend to come closer at higher frequency. Thus, since the intrinsic loss angles $\phi_\text{bulk}$ and $\phi_\text{shear}$ are continuous functions of the frequency, one should expect from Eq.~(\ref{eq:bulkshear_simple}) to see a similar trend in $\phi_c$, i.e. a relative narrowing of the differences between mode families with the increasing frequency. But existing silica loss data~\cite{granata2020amorphous} do not follow this trend and visibly show the opposite. In particular, the $(0,n)$ modes, which display their largest strain at the coating edge, tends to deviate more visibly from others at higher frequency. This clearly points to the existence of excess losses at the coating edge, which is expectedly non-uniform due to a variety of issues arising during deposition such as spill-off, tapering due to the sample holder, or lack of adhesion arising from imperfect surface polishing.

\begin{figure}
	\centering
	\includegraphics[width=0.8\columnwidth]{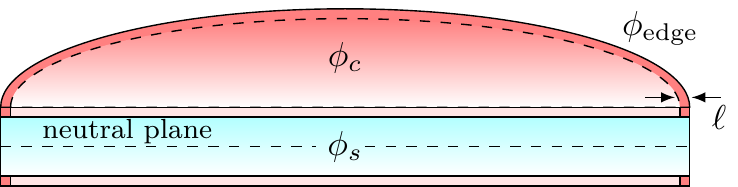}
	\caption{Cross-section of the disk, showing the substrate (blue), the bulk of the coating (light red), and the coating edge (deep red).}
	\label{fig:coating_barrel}
\end{figure}
In order to take these excess edge losses into account, we further divide the coating of a disk of radius $R$, as sketched in Fig.~\ref{fig:coating_barrel}, into a bulk region that extends over all radial distances $r<R-\ell$ and an outer annular ring of width $\ell$, i.e. corresponding to $r\in[R-\ell,R]$. The width $\ell$ is expected to be smaller than the coating thickness or at least of the same order since this is the only physical length in the problem. Furthermore, we assume the outer annular region to present a different loss angle than the rest of the disk. Applying again Eq.~(\ref{Phi_tot_generic}), while further separating the bulk and shear energies, the total loss angle now reads:
\begin{equation}
\begin{split}
	\phi_c &= (D_{\rm bulk}-D^{\rm edge}_{\rm bulk})\phi_{\text{\rm bulk}}+(D_{\rm shear}-D^{\rm edge}_{\rm shear})\phi_{\text{\rm shear}}\\
	&\qquad+ D^{\rm edge}_{\rm bulk}\phi^{\rm edge}_{\rm bulk}+D^{\rm edge}_{\rm shear}\phi^{\rm edge}_{\rm shear} \medspace. 
\end{split}
\label{eq:phiCinitial}
\end{equation}
Here $\phi_c$ is the total film loss angle, which can be experimentally accessed using equation~(\ref{eq:phiCbase}); $\phi_{\text{\rm bulk}}$ and $\phi_{\text{\rm shear}}$ are the intrinsic loss angles one seeks to measure to characterize the film; $D^{\rm edge}_{\rm bulk}$ and $D^{\rm edge}_{\rm shear}$ are the edge dilution factors for the bulk and shear elastic energies; and $\phi_{\rm bulk}^{\rm edge}$ and $\phi_{shear}^{\rm edge}$ are loss angles accounting for bulk and shear losses in the outer annular region.

We are exclusively concerned by situations in which the edge is small compared with the disk radius, i.e., $\ell\ll R$. In such cases, as we detail in Appendix~\ref{app:Dedge} (which extends a calculation presented in Ref.~\cite{cagnoli2017mode}), we find that the edge dilution factors are of the form:
\begin{equation}
    \begin{split}
        D^{\rm edge}_{\rm bulk}&=\frac{\ell}{R}\varepsilon_{\rm bulk}\\
        D^{\rm edge}_{\rm shear}&=\frac{\ell}{R}\varepsilon_{\rm shear}
    \end{split}
\end{equation}
where $\varepsilon_{\rm bulk}$ and $\varepsilon_{\rm shear}$ are dilution factor densities. The total film loss angle can then be recast as:
\begin{equation}
\begin{split}
    \phi_c &= 
	D_{\rm bulk}\phi_{\rm bulk} + D_{\rm shear}\phi_{\rm shear} \\
    	&\qquad +\frac{\ell}{R}\varepsilon_{\rm bulk}\delta\phi^{\rm edge}_{\rm bulk}
    	+\frac{\ell}{R}\varepsilon_{\rm shear}\delta\phi^{\rm edge}_{\rm shear}\medspace, 
\end{split}
	\label{eq:phiC}
\end{equation}
where $\delta\phi^{\rm edge}_{\rm bulk}=\phi^{\rm edge}_{\rm bulk}-\phi_{\rm bulk}$ and $\delta\phi^{\rm edge}_{\rm shear}=\phi^{\rm edge}_{\rm shear}-\phi_{\rm shear}$ are excess losses at the edge compared with the bulk values. 

\begin{figure}
	\centering
	\includegraphics[width=0.8\columnwidth]{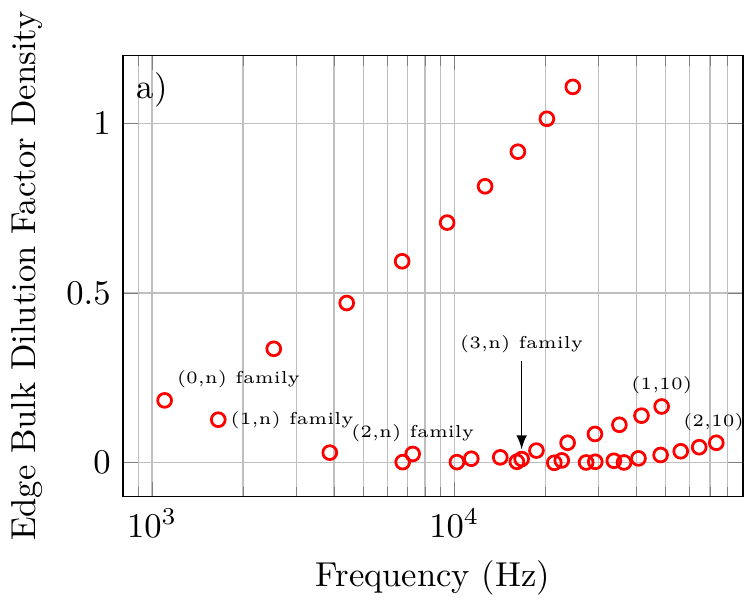}\quad
	\includegraphics[width=0.8\columnwidth]{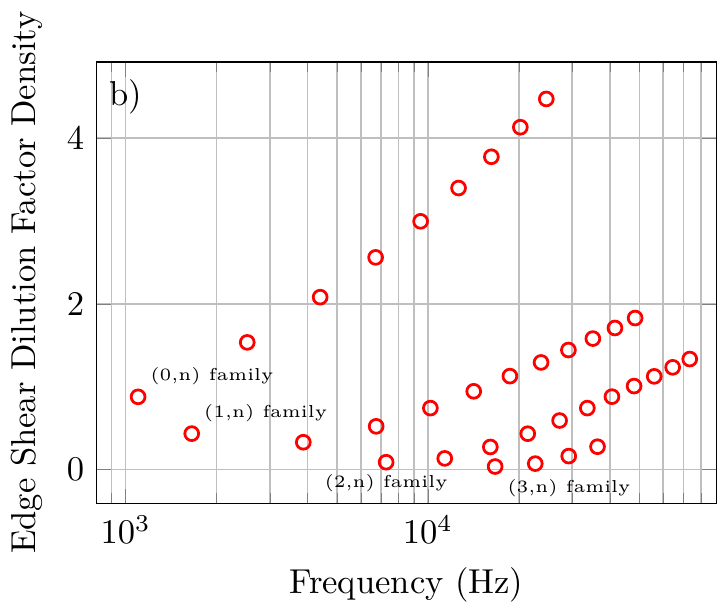}\quad
	\includegraphics[width=0.8\columnwidth]{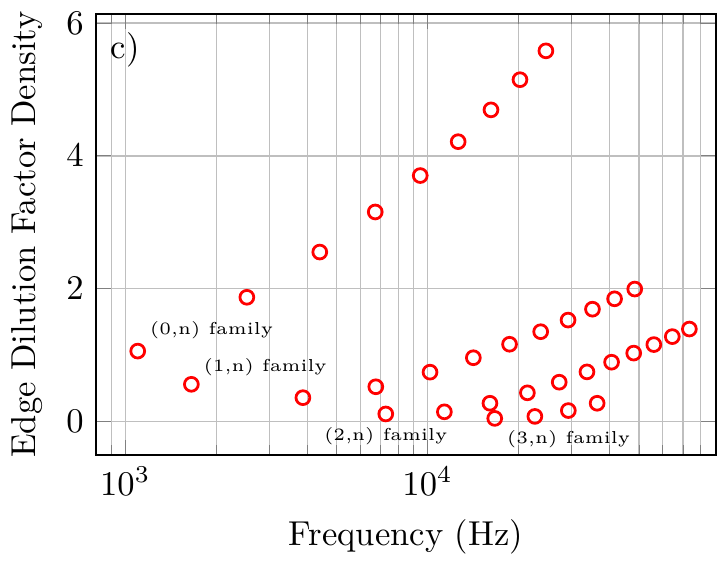}
	\caption{
	a) Bulk dilution factor density $\varepsilon_{\rm bulk}$, b) Shear dilution factor density $\varepsilon_{\rm shear}$ evaluated at the edge of a disk-shaped resonator and c) Total edge dilution factor density $\varepsilon = \varepsilon_{\rm bulk}+\varepsilon_{\rm shear}$. The frequency modes are those of a 75 mm, 1~mm thick, silica disk. For these numbers a Poisson's ratio of 0.16 for the mode shape $w(r,\theta)$ and 0.16 for the coating have been considered.}
	\label{fig:Dbulkshearedge}
\end{figure}

The values of $\varepsilon_{\rm bulk}$ and $\varepsilon_{\rm shear}$ are displayed in Appendix~\ref{app:Dedge} and in Table~\ref{tab:dilution} for various modes and parameters. To illustrate our discussion, we display them in Fig.~\ref{fig:Dbulkshearedge} using the same geometric parameters and Poisson ratio as in Figs.~\ref{fig:ansys} and~\ref{fig:Dbulkshear}. As expected, from our initial discussion of edge effects, they become more spread out at high frequency, because the elastic energy is then increasingly localized on the edge. This sharply contrasts with the bulk and shear dilution factors of Fig.~\ref{fig:Dbulkshear} where the values of $D_\text{bulk}$ and $D_\text{shear}$ for different $m$ are coming closer with the increasing frequency. Thus a rule-of-thumb guess for the importance of edge effects consists in considering the spread in loss angle values between modes of different azimuthal numbers: if it increases, or does not strongly reduce with the increasing frequency, this is a strong indication for the presence of an edge effect.

\subsection{Signature of an incomplete coating}

A particular type of edge effect arises when the coating deposition does not extend to the edge of the disk. This occurs in some deposition systems in which the sample holder masks the disk edge during deposition, leading to the formation of an coating-free annulus of width $\ell$ on the rim of the disk. As $\ell$ is typically of the order of few tenths of a millimetre, the condition $\ell\ll R$ holds and the coating loss angle can be written as:
\begin{equation}
\phi_c = \left(D_\text{bulk}-\frac{\ell}{R}\varepsilon_\text{bulk}^{\rm edge}\right)\phi_{\rm bulk} +\left(D_\text{shear}-\frac{\ell}{R}\varepsilon_\text{shear}^{\rm edge}\right)\phi_{\rm shear}
\end{equation}
This is just the same equation as Eq.~(\ref{eq:phiC}), yet with excess edge loss angles that are \emph{negative} and exactly opposite to the bulk values:
\begin{equation}\label{eq:noedge:dphi}
\begin{split}
    \delta\phi^{\rm edge}_{\rm bulk}&=-\phi_{\rm bulk}\\
    \delta\phi^{\rm edge}_{\rm shear}&=-\phi_{\rm shear}\medspace.
\end{split}
\end{equation}

\section{Case study: a power law material with edge effect}

In order to illustrate how edge inhomogeneities affect GeNS measurements, let us consider an idealized coating material obeying exactly the TLS prediction, i.e., displaying bulk and shear loss angles, $\phi_\text{bulk}$ and $\phi_\text{shear}$ that are power law functions of the frequency. Under these assumptions, equation~(\ref{eq:phiC}) can be written:
\begin{equation}
\begin{split}
\phi_c &=A_1\, \left(\frac{f}{10\text{kHz}}\right)^{\alpha_1}D_\text{bulk} \\ 
&\qquad+A_2\, \left(\frac{f}{10\text{kHz}}\right)^{\alpha_2}D_\text{shear}\\
&\qquad\quad+\frac{\ell}{R}\varepsilon_{\rm bulk}\delta\phi^{\rm edge}_{\rm bulk}
+\frac{\ell}{R}\varepsilon_{\rm shear}\delta\phi^{\rm edge}_{\rm shear}\medspace, 
\end{split}
\label{eq:final}
\end{equation}
where we have redefined the coupling constants $A_1$ and $A_2$ to normalize the frequency by 10kHz, a typical scale in the range at which data are obtained. We will refer to equation~(\ref{eq:final}) as the BSE model (for bulk, shear, and edge).  

Here, in order to better illustrate edge-induced artifacts, we will additionally use the conservative assumption that the same two-level systems couple to both bulk and shear strains, in which case $\alpha_1=\alpha_2=\alpha$. In addition, we will consider for simplicity that the excess edge loss angles are frequency-independent and identical, $\delta\phi_\text{bulk}^\text{edge}=\delta\phi_\text{shear}^\text{edge}=\delta\phi^\text{edge}$, a constant, in which case, the two edge terms
$\frac{\ell}{R}\varepsilon_{\rm bulk}\delta\phi^{\rm edge}_{\rm bulk}+\frac{\ell}{R}\varepsilon_{\rm shear}\delta\phi^{\rm edge}_{\rm shear}=\frac{\ell}{R}\varepsilon\delta\phi^\text{edge}$ with $\varepsilon=\varepsilon_{\rm bulk}+\varepsilon_{\rm shear}$. Thus restricted, the BSE model comprises only four parameters: $A_1$, $A_2$, $\alpha$, and $\delta\phi^\text{edge}$.

An experimentalist studying this material may be tempted to fit $\phi_c$ using equation~(\ref{eq:bulkshear}), which we henceforth call the BS (bulk and shear) model. This model overlooks edge effects and uses different power law exponents for bulk and shear losses, thus comprising four fitting parameters, $A_1$, $A_2$, $\alpha_1$, and $\alpha_2$.

\begin{figure*}
	\centering
	\includegraphics[width=1.55\columnwidth]{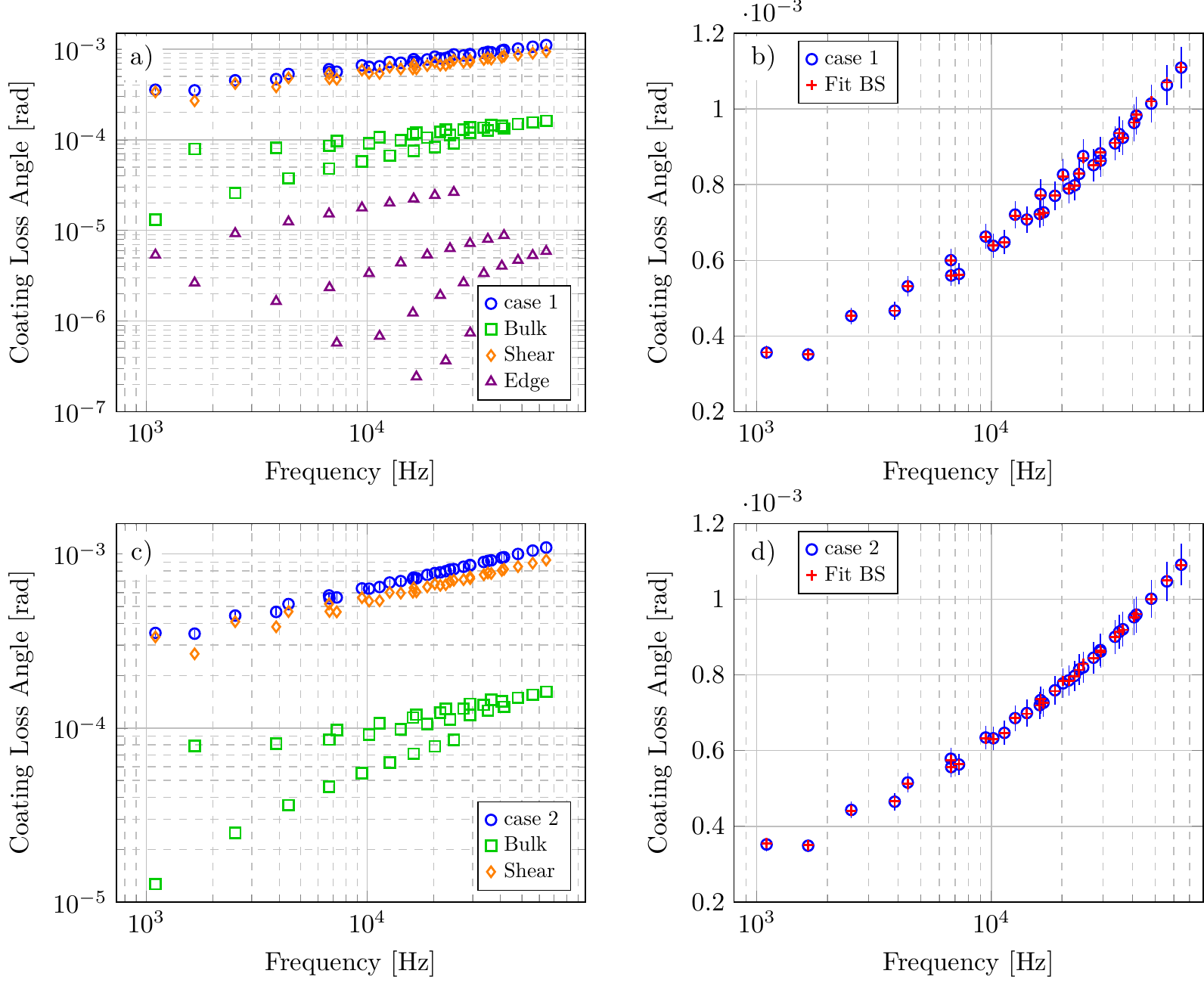}
	\caption{BS fits of BSE data for the low frequency modes of a $\diameter$ 75 mm, 1~mm thick silica disk-shaped resonator with a $\nu=0.16$ coating. The parameters are listed in table \ref{tab:simulations} and the dilution factors in table \ref{tab:dilution}. Top: a complete coating with excess edge loss; bottom, an incomplete coating. Left panels: decomposition of the BSE data (blue $\Circle$) into bulk (green $\square$), shear (orange $\Diamond$) and the edge (violet $\triangle$, top) terms. Right panels: comparison of the BS fits with the BSE data tainted with a 5\% error. The BS model fits the BSE data remarkably well, yet with erroneous estimates of the material parameters, as shown in table \ref{tab:simulations}.}
	\label{fig:simulations}
\end{figure*}

In order to simulate the experimental process, we use the BSE model to generate a complete set of $\phi_c$ values for the low frequency modes of a 75 mm silica disk. Specifically, we assume the coating has a Poisson ratio $\nu=0.16$, and an exponent $\alpha=0.3$, a value that was found in silica at 100K~\cite{PhysRevB.64.064207,TRAVASSO2009268}, and which is hence highly plausible for coatings in a low temperature range. In addition, as seen on Fig.~\ref{fig:simulations}a, we targeted a total film loss angle in the range of $10^{-4}$, which is typical of low noise applications, while making sure, for the sake of the illustration, that the edge term $\ell\varepsilon \delta\phi^\text{edge}$ remained at least 3 times smaller than the other contributions for each frequency.

\begin{table*}
	\caption{\label{tab:simulations} Parameters used to generate the data and fit results with the mean squared error (MSE), obtained using equation \ref{eq:bulkshear} (BS) and the Levemberg-Marquadt minimisation algorithm. The elements of the correlation matrix $\rho_{ij}$ have been obtained by the elements of the covariance matrix $C_{ij}$ up to 1 standard deviation from the minimum of the $\chi^2$ function, $\rho_{ij}=C_{ij}/\sqrt{C_{ii}C_{jj}}$.
	They show that the bulk and shear parameters are strongly anti-correlated, presumably because the associated dilution are complementary to 1. The correlation is higher between either $A_1$ and $A_2$ or $\alpha_1$ and $\alpha_2$, which significantly increases the uncertainties on these fitted values.}
	\begin{ruledtabular}
		\begin{tabular}{lcccccc}
			 & $\ell\phi^{\rm bulk}_{\rm edge}$ $(\times10^{-7})$ & $\ell\phi^{\rm shear}_{\rm edge}$ $(\times10^{-7})$  & $A_1$ $(\times10^{-4})$ &   $\alpha_1$ $(\times10^{-1})$  &   $A_2$ $(\times10^{-4})$  &   $\alpha_2$  $(\times10^{-1})$     \\
			\multicolumn{7}{l}{\textbf{Case 1}}\\
			Parameters:	& 2 & 2 & 4  & 3 & 7 & 3   \\
			Fit BS: \scriptsize{MSE = 0.003}&- &-  & $2.80\pm0.07$  & $1.26\pm0.16$  & $7.408\pm0.017$ & $3.181\pm0.014$  \\
			\addlinespace[1ex]
			\colrule
			\addlinespace[1ex]
			\multicolumn{7}{l}{\textbf{Case 2}}\\
			Parameters: ($l=0.5$ mm)	& 2 & 2 & 4  & 3 & 7 & 3   \\
			Fit BS: \scriptsize{MSE = 0.006}  & -& -  & $4.85\pm0.08$ & $3.98\pm0.11$  & $6.70\pm0.002$ & $2.75\pm0.03$  \\
			\addlinespace[1ex]
			\colrule
			\addlinespace[1ex]
			\multicolumn{7}{c}{\textbf{Correlation Matrix}}\\
			\addlinespace[1ex]
			\multicolumn{3}{c}{
				$\text{case 1}=\begin{pmatrix}  
				& A_1 & \alpha_1 & A_2 & \alpha_2 \\ 
				A_1	&	1.000	&	0.806	&	-0.969	&	-0.706	\\
                \alpha_1	&	0.806	&	1.000	&	-0.765	&	-0.926	\\
                A_2	&	-0.969	&	-0.765	&	1.000	&	0.631	\\
                \alpha_2	&	-0.706	&	-0.926	&	0.631	&	1.000	\\
				\end{pmatrix}$} &
			\multicolumn{4}{c}{
				$\text{case 2}=\begin{pmatrix}   
				& A_1 & \alpha_1 & A_2 & \alpha_2 \\ 
				A_1	&	1.000	&	0.410	&	-0.959	&	-0.618	\\
                \alpha_1	&	0.410	&	1.000	&	-0.520	&	-0.901	\\
                A_2	&	-0.959	&	-0.520	&	1.000	&	0.665	\\
                \alpha_2	&	-0.618	&	-0.901	&	0.665	&	1.000	\\
				\end{pmatrix}$} \\
			\addlinespace[2ex]
		\end{tabular}
	\end{ruledtabular}
\end{table*}

After adding a 5\% error on every data point, we fit these data using the BS model, equation~(\ref{eq:bulkshear}): the BS fit, displayed in Fig.~\ref{fig:simulations}b, is remarkably convincing, although erroneous by construction. The parameters used to generate the data and the fit results are listed in table~\ref{tab:simulations}. We see from their comparison that the fitted material properties ($A_1$, $A_2$, $\alpha_1$ and $\alpha_2$) are grossly erroneous even though the edge loss contribution is quite small compared with all other terms. It is especially striking that the fitted power law exponents present distinct values, while the idealized BSE material does not. For the sake of completeness, we examined (not shown) cases when $\alpha_1$ and $\alpha_2$ are distinct in the BSE model, and found that the fitted exponents (as well as $A_1$ and $A_2$) are again systematically grossly erroneous. Moreover, when the edge loss contribution increases and approaches the bulk and shear contributions, we also found that the fitted exponents may become negative, which is unphysical.

It follows from these observations that presence of a very small edge effect induces a severe error on the intrinsic loss properties of the film as deduced from BS fitting.\\ 

Let us now use our test scenario in order to illustrate how an incomplete coating also distorts the measurement of material properties. A case of interest for most deposition methods when the sample holder covers the edge of the disk during deposition. We thus consider that the coating is just the same material as in the above example, and is deposited on an identical 3'' disk with a $\nu=0.16$ Poisson's ratio. However, we now assume that a the edge effects arises solely from the absence of coating over a $\ell=0.5$mm outer rim. The coating loss angle hence is described by the following equation:
\begin{equation}
\begin{split}
\phi_c &=A_1\, \left(\frac{f}{10\text{kHz}}\right)^{\alpha_1}\left(D_\text{bulk}- \frac{\ell}{R}\varepsilon_{\rm bulk}\right)\\ 
&\qquad+A_2\, \left(\frac{f}{10\text{kHz}}\right)^{\alpha_2}\left(D_\text{shear}- \frac{\ell}{R}\varepsilon_{\rm shear}\right)
\end{split}
\label{eq:phiNoEdge}
\end{equation}
which is identical to Eq.~(\ref{eq:final}), yet for $\delta\phi^{\rm edge}_{\rm bulk}$ and $\delta\phi^{\rm edge}_{\rm shear}$ values that are opposite to their bulk counterpart. 

As before, we fit the resulting loss angle data using the BS model so as to mimic an erroneous analysis. The fit, which is displayed in Fig.~\ref{fig:simulations}d looks excellent with an MSE of 0.006. While the fitted values of $A_1$ and $A_2$ remain comparable with the actual values, the exponent $\alpha_1$ is tainted with an error of over 30\% and more strikingly the exponents appear to be different while they actually are not.

Therefore, to access material properties, one should worry about the possible tapering of the coating at the edge. When the coating is incomplete, one should fit the data using an equation of the form~(\ref{eq:phiNoEdge}) with a precise measurement of the width $\ell$ of the uncoated region. One may further need to consider a model comprising both the tapering and an edge contribution arising from the outer rim of the coated area.

\section{Analysis of experimental data}

Let us now assess the relevance of the edge effect to the analysis of real data. For this purpose, we consider room temperature loss angle measurements for tantala (\ce{Ta2O5}) and silica (\ce{SiO2}) coatings deposited under the same conditions as those used for the manufacture of gravitational-waves detectors~\cite{granata2020amorphous}, namely by ion-beam sputtering at the Laboratoire des Matériaux Avancés. In both cases, to avoid additional stress coming from the mismatch of mechanical properties between substrate and coating, deposition was performed on both sides of a silica disk-shaped resonator ($\diameter$ 50 mm and 75 mm and 1 mm thick).

The coating loss angles data were fitted with either the BS [Eq.~(\ref{eq:bulkshear})] or the BSE model [Eq.~(\ref{eq:final})]. For the BSE fit, we tried to use different power law exponents or distinct excess edge loss angles for the bulk and shear components. However, we found that the MSE (which depends on the number of fitted parameters) is smaller when using a unique power law exponent and a unique excess loss angle. Moreover, we found that the power law exponents was so small that it was compatible with zero. This is consistent with the observation by Travasso~\textit{et al.}~\cite{PhysRevB.64.064207,TRAVASSO2009268} that $\alpha$ grows linearly with $T$, as predicted by the TLS theory, only up to about 120K, but decreases significantly with the further increase of temperature. This led us to use a reduced BSE model in which $\alpha_1=\alpha_2=0$ and with a single excess edge contribution, hence comprising overall just three fitting parameters:
\begin{equation}
\phi_c =A_1\,D_\text{bulk} 
+A_2\,D_\text{shear}
+\frac{\ell}{R}\varepsilon\delta\phi^{\rm edge}
\label{eq:BSE-r}
\end{equation}
Subsequently, we refer to this model as the reduced BSE, or BSE-r for short.

\begin{figure*}
	\centering
	\includegraphics[width=1.4\columnwidth]{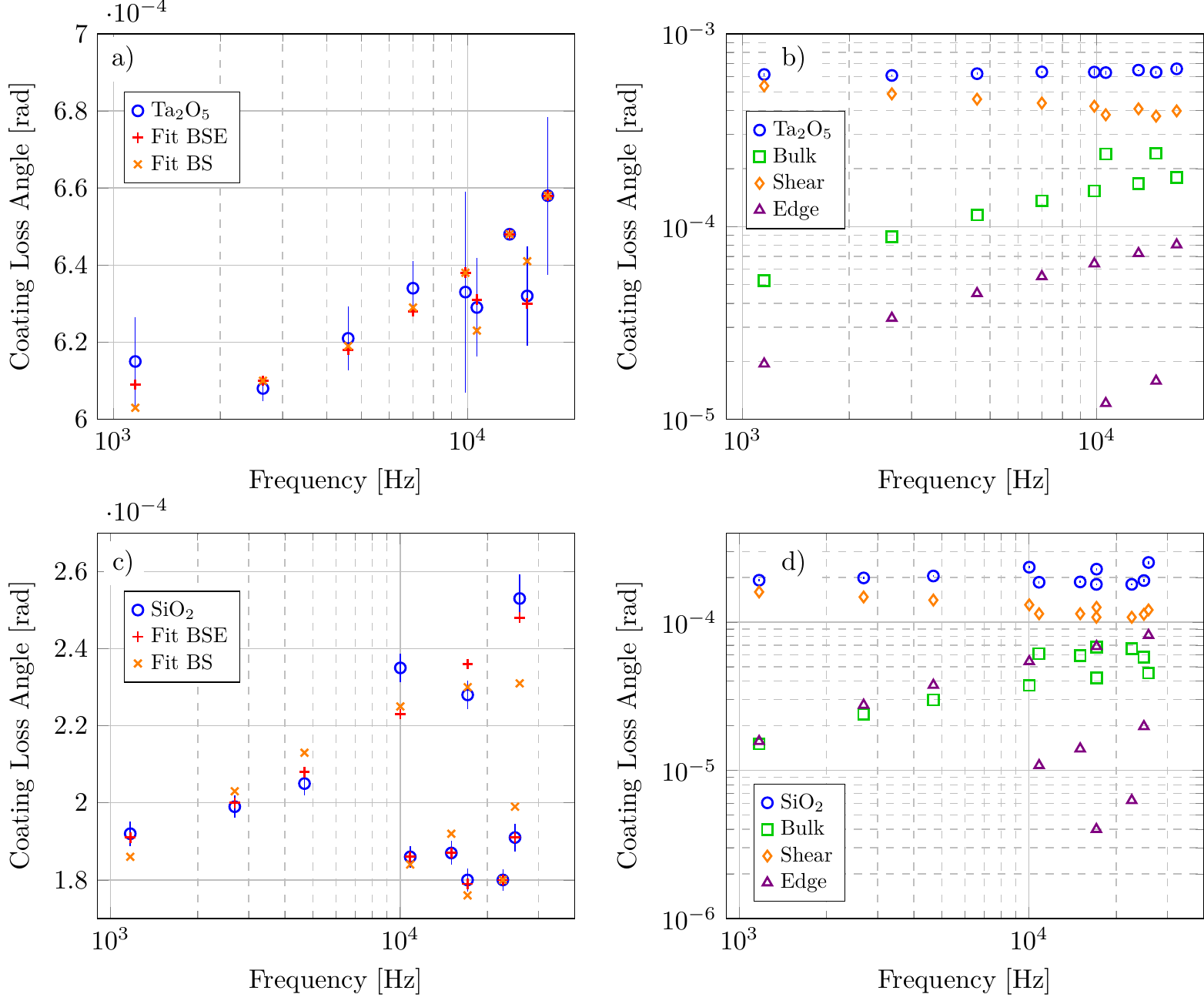}
	\caption{Analysis of loss angle data for \ce{Ta2O5} (top) and \ce{SiO2} (bottom) coatings. Left panels: fits of the data (blue) with the BS [orange crosses, Eq.~(\ref{eq:bulkshear})] or the BSE-r [red plus symbols, Eq.~(\ref{eq:final})] model. Right panels: the bulk (green squares), shear (orange diamonds), and edge (purple triangles) contributions as deduced by the BSE-r fit.}
	\label{fig:Data}
\end{figure*}

We fit both \ce{Ta2O5} and \ce{SiO2} coatings data using either the BS or BSE-r models. The fit parameters are reported in table~\ref{tab:Data} and the fits in the left panels of Figure~\ref{fig:Data}. The right panels of this figure present a decomposition of the loss angle into the bulk, shear, and edge contributions (each of which is weighted by the proper dilution factor), as deduced from the BSE-r fit.

\begin{table*}
	\caption{\label{tab:Data}Fit results obtained by the Levemberg-Marquadt algorithm by analysing different coatings using equation \ref{eq:bulkshear} and \ref{eq:final}. The elements of the correlation matrix $\rho_{ij}$ have been obtained by the elements of the covariance matrix $C_{ij}$ up to 1 standard deviation from the minimum of the $\chi^2$ function, $\rho_{ij}=C_{ij}/\sqrt{C_{ii}C_{jj}}$. The best fits using the full BSE model gives exponent values compatible with 0, which motivates using the reduced (BSE-r, see text) model.}
	\begin{ruledtabular}
		\begin{tabular}{lcccccc}
			\multicolumn{7}{c}{\textbf{\textbf{\ce{Ta2O5}}}}\\
			Parameters: & $\ell\phi^{\rm bulk}_{\rm edge}$ $(\times10^{-7})$ & $\ell\phi^{\rm shear}_{\rm edge}$ $(\times10^{-7})$  & $A_1$ $(\times10^{-4})$ &   $\alpha_1$ $(\times10^{-1})$  &   $A_2$ $(\times10^{-4})$  &   $\alpha_2$  $(\times10^{-1})$     \\
			Fit BS: \scriptsize{MSE = 0.60} & - & - & $4.3\pm0.9$  & $3\pm2$ & $6.74\pm0.16$ & $0.31\pm0.13$   \\
			Fit BSE-r: \scriptsize{MSE = 0.72} & $4.4\pm0.6$ & $=\ell\phi^{\rm bulk}_{\rm edge}$ & $8.1\pm0.6$ & -  & $5.67\pm0.07$ & -  \\
			\addlinespace[1ex]
			\multicolumn{7}{c}{Correlation Matrix}\\
			\multicolumn{3}{c}{Fit BS} & \multicolumn{4}{c}{Fit BSE-r} \\
			\multicolumn{3}{c}{
				$\begin{pmatrix}  
					& A_1 & \alpha_1 & A_2 & \alpha_2 \\ 
					A1	&	1.000	&	-0.405	&	-0.988	&	-0.500	\\
					B1	&	-0.405	&	1.000	&	0.273	&	-0.562	\\
					A2	&	-0.988	&	0.273	&	1.000	&	0.608	\\
					B2	&	-0.500	&	-0.562	&	0.608	&	1.000	\\
				\end{pmatrix}$} &
			\multicolumn{4}{c}{
				$\begin{pmatrix}  
					& A_1 & A_2 & \ell\phi^\text{edge}\\ 
					A_1	&	1.000	&	-0.688	&	-0.706	\\
                    A_2	&	-0.688	&	1.000	&	-0.016	\\
                    \ell\phi^\text{edge}	&	-0.706	&	-0.016	&	1.000	\\
				\end{pmatrix}$} \\ 
			\addlinespace[1ex]
			\colrule
			\addlinespace[1ex]
			\multicolumn{7}{c}{\textbf{\ce{SiO2}}}\\
			Parameters: & $\ell\phi^{\rm bulk}_{\rm edge}$ $(\times10^{-7})$ & $\ell\phi^{\rm shear}_{\rm edge}$ $(\times10^{-7})$  & $A_1$ $(\times10^{-4})$ &   $\alpha_1$ $(\times10^{-1})$  &   $A_2$ $(\times10^{-4})$  &   $\alpha_2$  $(\times10^{-1})$     \\
			Fit BS: \scriptsize{MSE = 6.18} & -&- & -$1.0\pm0.5$  & $8\pm3$ & $2.92\pm0.13$ & $1.7\pm0.3$   \\
			Fit BSE-r: \scriptsize{MSE = 2.36} & $4.9\pm0.5$ & $=\ell\phi^{\rm bulk}_{\rm edge}$ & $1.72\pm0.17$  & -  & $1.77\pm0.06$ & -  \\
			Fit BSE-r': \scriptsize{MSE = 2.11} & $4.9\pm0.4$ & $=\ell\phi^{\rm bulk}_{\rm edge}$ & $1.76\pm0.02$ & -  & $=A_1$  & -   \\
			\addlinespace[1ex]
			\multicolumn{7}{c}{Correlation Matrix}\\
			\multicolumn{3}{c}{Fit BS} & \multicolumn{2}{c}{Fit BSE-r} & \multicolumn{2}{c}{Fit BSE-r'}\\
			\multicolumn{3}{c}{
				$\begin{pmatrix}  
						&	A1	&	B1	&	A2	&	B2	\\
				A1	&	1.000	&	0.734	&	-0.914	&	-0.630	\\
				B1	&	0.734	&	1.000	&	-0.455	&	-0.044	\\
				A2	&	-0.914	&	-0.455	&	1.000	&	0.766	\\
				B2	&	-0.630	&	-0.044	&	0.766	&	1.000	\\
				\end{pmatrix}$} &
			\multicolumn{2}{c}{
				$\begin{pmatrix}  
					&	A_1	&	A_2	&	\ell\phi^\text{edge}	\\
                A_1	&	1.000	&	-0.846	&	0.350	\\
                A_2	&	-0.846	&	1.000	&	-0.692	\\
                \ell\phi^\text{edge}	&	0.350	&	-0.692	&	1.000	\\
				\end{pmatrix}$} & 
				\multicolumn{2}{c}{
				$\begin{pmatrix}  
					&	A	&	\ell\phi^\text{edge}	\\
				A	&	1.000	&	-0.766		\\
				\ell\phi^\text{edge}	&	-0.766	&	1.000			\\
				\end{pmatrix}$} \\ 
			\addlinespace[1ex]
		\end{tabular}
	\end{ruledtabular}
\end{table*}		

For \ce{Ta2O5}, the BSE-r fit has only a slightly larger MSE that the BS one, yet within a range compatible with irreducible experimental errors. As we already said, the BS fit is problematic for two reasons. First, because the TLS model is not demonstrated to hold up to room temperature. Second, because even if it extended up to these conditions, one should expect both bulk and shear exponents to display a unique value, whereas the fit only works for widely different values of $\alpha_1$ and $\alpha_2$. For these reasons, the reduced BSE fit is far more satisfactory. It hence supports that: (i) the bulk and shear loss angles are essentially frequency-independent on the investigated range (1 kHz to 40 kHz); (ii) the edge contribution is systematically smaller than the bulk and shear ones; (iii) strikingly, this small edge contributions suffices to mislead the BS fit into predicting spurious frequency dependencies of the bulk and shear loss angles.

For \ce{SiO2}, the BS fit is grossly inconsistent since it yields an $A_1$ value which in negative. Instead, the BSE-r fit works remarkably well and displays an MSE which is three times smaller than the BS one. Compared with the \ce{Ta2O5} data, the edge term has a comparable magnitude, ranging between $\num{1e-5}$ and $\num{1e-4}$, while the shear and bulk contributions are smaller. This results in a stronger relative importance of the edge contribution, which may be larger than some bulk contributions (compare the upper triangle and squares in the bottom right panel of Fig.~\ref{fig:Data}). This effect would seem to explain why the BS fit works so poorly for \ce{SiO2} coatings.

Finally, when fitting \ce{SiO2} coatings data, we noted that the BSE-r fit produces $A_1$ and $A_2$ values that are close to one another. This led us to perform, tentatively, an even more reduced fit, which we call BSE-r', in which it is assumed that $A_1=A_2$. In such a case, the loss angle reduces to the remarkably simple expression:
\begin{equation}
\phi_c =A +\frac{\ell}{R}\varepsilon\delta\phi^{\rm edge}
\label{eq:BSE-r'}
\end{equation}
with $A\equiv A_1=A_2$ being a constant. This fit works remarkably well, and displays a smaller MSE than the BSE-r one. It predicts that the frequency dependence of $\phi_c$ comes solely from the edge term, more specifically, from the edge dilution factor density $\varepsilon$. This is consistent with the observation that, in silica coatings, the $\phi_c$ values for different mode families (different radial numbers) are strongly splitted at higher frequency, a signature of excess edge loss contributions.

The reason why $A_1$ and $A_2$ should be identical on fundamental grounds remains uncertain. It would seem to suggest that, in silica, the bulk and shear strains are strongly coupled and disperse energy via a unique channel, an idea that is certainly worth exploring in the future.

\section{Conclusion}

In this work, we have investigated the consequences of a systematic error arising from edge effects on the measurement of coating loss angle in disk-shaped resonators.
Excess edge losses may arise from a variety of reasons such as, to name a few: coating thickness non-uniformity at the edge, coating spill-off during the deposition, tapering due to the sample holder during deposition, or coating deposition on an unpolished surface and an associated lack of adhesion near the edge. They are hence expected to be widespread and should be systematically assessed when analyzing experimental data.

By producing artificial loss data, we showed that this excess edge loss contribution could introduce a significant error in the fitted values of the bulk and shear loss angles. This idea was confirmed by the analysis of experimental data for \ce{Ta2O5} and \ce{SiO2} coatings. The BS model, indeed, appeared at best suspicious ($\alpha_1\ne\alpha_2$) and at worst fully inconsistent. Meanwhile, the BSE-r model introduces a very limited set of assumptions and fewer parameters, while yielding a mean-square error that is comparable (\ce{Ta2O5}) or much smaller (\ce{SiO2}).

These results unambiguously support that coating loss angle measurements are affected by an excess edge contribution which, although small in value, dramatically affects the fitting procedure and may lead to the attribution of erroneous values to the bulk and shear loss angles, which are the key intrinsic material properties targeted by these  measurements. In particular, it now appears that, in room temperature conditions, over the accessible frequency range, any frequency-dependence of $\phi_\text{bulk}$ and $\phi_\text{shear}$ is too small to be measurable using GeNS data.

\section*{Acknowledgement}
This work benefited from the support of the project ViSIONs ANR-18-CE08-0023-03 of the French National Research Agency (ANR).

\appendix

\section{Dilution factor coating/substrate}
\label{app:D}
When the disk is free to oscillate in one of its resonant modes, its elastic energy must be equal to the kinetic energy,
\begin{equation}
	K=\frac{1}{2}\rho\omega^2\int w^2(r,\theta)dzrdrd\theta\medspace,
	\label{eq:Kinenergy}
\end{equation}
where $\rho$ is the mass density, $w(r,\theta)$ is the maximum off-plane displacement that represents the mode shape in circular coordinates and $\omega$ is the angular frequency. 
By solving the integral for the bare substrate thickness $h$ along $z$, we obtain
\begin{equation}
	K_s=\frac{1}{2}\rho_sh\omega_s^2\int w_s^2(r,\theta)rdrd\theta\medspace,
	\label{eq:KinenergySub}
\end{equation}
whereas for a sample coated on both sides, with a coating thickness $c$,
\begin{equation}
	K_\text{tot}=\frac{1}{2}\rho_\text{tot}(h+2c)\omega_\text{tot}^2\int w_\text{tot}^2(r,\theta)rdrd\theta\medspace.
	\label{eq:KinenergyTot}
\end{equation}
Since the kinetic energy must be equal to the elastic energy, it is possible to write
\begin{equation}
	\begin{dcases}
		\frac{1}{2}\rho_sh\omega_s^2\int w_s^2(r,\theta)rdrd\theta=E_s\medspace,\\
		\frac{1}{2}\rho_\text{tot}(h+2c)\omega_\text{tot}^2\int w_\text{tot}^2(r,\theta)rdrd\theta=E_\text{tot}\medspace.
		\label{eq:Kinenergycases}
	\end{dcases}
\end{equation}
The integrals in equations (\ref{eq:Kinenergycases}) depend on the mode shape $w(r,\theta)$ of the disk (see appendix \ref{app:bulkshear}).
The elastic energy of the substrate $E_s$ and of the coated sample $E_\text{tot}$ can be evaluated analytically~\cite{leissa1969vibration}.

Since the coating is substantially small compared to the substrate, it reasonable to assume that the mode shape of the coated disk $w_\text{tot}$ is equal to that of the bare disk $w_s$. This implies also that the neutral plane -- the imaginary surface that has zero deformation at all times during the oscillation -- remains in the same position. Under this assumption, dividing each members of equations (\ref{eq:Kinenergycases}) we obtain
\begin{equation}
	\frac{\rho_sh}{\rho_\text{tot}(h+2c)}\left(\frac{\omega_s}{\omega_\text{tot}}\right)^2=\frac{E_s}{E_\text{tot}}\medspace.
\end{equation}
Considering that $E_\text{tot}=E_s+E_c$ and that $A\rho_sh=m_s$ and $A\rho_\text{tot}(h+2c)=m_\text{tot}$, where $A$ is the area of the disk and $m_s$, $m_\text{tot}$ the mass of the disk before and after the coating deposition, respectively, we obtain
\begin{equation}
	D=1-\frac{m_s}{m_\text{tot}}\left(\frac{\omega_s}{\omega_\text{tot}}\right)^2\medspace.
	\label{eq:dilution}
\end{equation}
where $D=E_c/E_\text{tot}$ is the dilution factor. 
Equation \ref{eq:dilution} is very important because it gives the dilution factor as a function of measurable parameters, without any prior knowledge of coating elastic constants or thickness. The masses and frequencies after coating deposition are significantly different from those of the bare substrates, so that they can be measured with an analytical balance and a GeNS system, respectively. However, the validity of equation \ref{eq:dilution} is based on the homogeneity of the coating on the entire surface of the substrate. Since the resonant frequency depend on the disk curvature, the same coating must be deposited on both sides of the substrate, avoiding additional stress and bending of the sample~\cite{GranataInternalFriction}. A work dedicated to the stress dependence of the mode frequencies will be submitted soon by some of the authors.

Once the substrate loss is subtracted from the measurements, the coating can be considered as an independent disk. Indeed, assuming that the coating dissipation pathways can be related to angles $\phi_1$, $\phi_2$,\ldots, the total loss angle reads
\begin{equation}
    \phi_\text{tot}=(1-D)\phi_s+D\left(D_1\phi_1+D_2\phi_2+\cdots\right)\medspace,
\end{equation}
with $D=E_c/E_\text{tot}$: we see that the coatings dilution factors are $D_i=E_i/E_c$ can be calculated by identifying the energies stored in the coating subdomains and degrees of freedom only.

\section{Dilution factor for bulk and shear}
\label{app:bulkshear}
A deformation can be decomposed in bulk, where the dilatation changes the volume of the sample, and shear, where the deformation does not affect the volume. Therefore, the free energy density can be written as~\cite{landau1986theory}
\begin{align}
	E &= \frac{1}{2}(\lambda + \frac{2}{3}\mu)u_{ll}^2+\mu(u_{ik}-\frac{1}{3}\delta_{ik}u_{ll})^2\\
	&= \frac{1}{2}Ku_{ll}^2+\mu(u_{ik}-\frac{1}{3}\delta_{ik}u_{ll})^2\medspace, \label{eq:Freeenergybulk}
\end{align}
where $u_{ik}$ are the elements of the strain tensor, $\lambda$, $\mu$ are the Lamé's coefficients and $K=\lambda + \frac{2}{3}\mu$ is the bulk modulus, which are related to the Young's modulus $Y$ and Poisson's ratio $\nu$ as follows
\begin{equation}
	Y=\frac{9K\mu}{3K+\mu}\medspace,\qquad\nu=\frac{1}{2}\frac{3K-2\mu}{(3K+\mu)}\medspace,
\end{equation}
so that
\begin{equation}
	K=\frac{1}{3}\frac{Y}{(1-2\nu)}\medspace.
\end{equation}
The first term of equation \ref{eq:Freeenergybulk} is the energy related to the bulk modulus and can be expressed in polar coordinates as
\begin{align}
	u_{rr}&= \frac{\partial u_r}{\partial r}\medspace,\\
	u_{\theta\theta}&= \frac{1}{r}\frac{\partial u_\theta}{\partial \theta}+\frac{ u_r}{r}\medspace,\\
	u_{zz}&= -\frac{\nu}{1-\nu}\left(u_{rr}+u_{\theta\theta}\right)\medspace,
\end{align}
the latter comes from the imposition of the condition $\sigma_{zz}=0$. Furthermore, $u_{r}$, $u_{\theta}$ and $u_{z}$ can be expressed by three off-plane displacement $w$ as in the following
\begin{align}
	u_{r}&= -z\frac{\partial w}{\partial r}\medspace,\\
	u_{\theta}&= -\frac{z}{r}\frac{\partial w}{\partial \theta}\medspace,\\
	u_{z}&= w\medspace.
\end{align}
The infinitesimal bulk energy is
\begin{align}
	dE_{\text{bulk}}&= \frac{1}{2}Ku_{ll}^2rdrd\theta dz \\
	&= \frac{1}{6}\frac{Y}{(1-2\nu)}\left(-z\frac{\partial^2 w}{\partial r^2}-\frac{z}{r^2}\frac{\partial^2 w}{\partial \theta^2}-\frac{z}{r}\frac{\partial w}{\partial r}\right)^2rdrd\theta dz \label{eq:Freeenergybulk_final_intra}\\
	&=\frac{Y}{6(1-2\nu)}(\nabla^2w)^2z^2rdrd\theta dz\medspace.
	\label{eq:Freeenergybulk_final}
\end{align}
The infinitesimal total elastic energy is instead~\cite{leissa1969vibration}
\begin{multline}
	dE=\frac{Y}{1+\nu}\biggl\{\frac{1}{2(1-\nu)}\left(\frac{\partial^2w}{\partial r^2}+\frac{1}{r}\frac{\partial w}{\partial r}+\frac{1}{r^2}\frac{\partial^2 w}{\partial\theta^2}\right)^2- \\
	+\biggl[\frac{\partial^2w}{\partial r^2}\left(\frac{1}{r}\frac{\partial w}{\partial r}+\frac{1}{r^2}\frac{\partial^2 w}{\partial\theta^2}\right)- \\
	+\left(\frac{\partial}{\partial r}\left(\frac{1}{r}\frac{\partial w}{\partial\theta}\right)\right)^2\biggr]\biggr\}z^2dzrdrd\theta\medspace.
	\label{eq:energydisk}
\end{multline}
From the equation of motion (not shown here) one can derive the expression for the off-plane displacement $w$
\begin{equation}
w(r,\theta) = A_{m,n}\left[J_n\left(\lambda_{m,n}\frac{r}{R}\right)+C_{m,n}I_n\left(\lambda_{m,n}\frac{r}{R}\right)\right]\cos(n\theta)\medspace,
\end{equation}
where $C_{m,n}$ and $\lambda_{m,n}$ are obtained by the boundary conditions, which in the case of free-edge vibrating plate correspond to the Kirchhoff-Kelvin boundary conditions at the edge of the disk~\cite{amabili1995natural}, $J_n$ and $I_n$ are the Bessel functions and the modified Bessel functions of the first type, respectively. Therefore, for each mode $(m,n)$ it is possible to integrate the infinitesimal energies \ref{eq:Freeenergybulk_final} and \ref{eq:energydisk}. The coefficient $A_{m,n}$ simplifies once the ratio is performed. All along this article we have considered the mode shapes $w_{m,n}$ as they come from a substrate of fused silica with $\nu=0.16$. The coating can have a different Poisson's ratio but its thickness is so small that in first approximation the mode shapes $w_{m,n}$ are fixed only by the substrate properties. Only exception at this approximation is when the coating Poisson's ratio is very close to 0.5. In that case, the coating bulk energy becomes dominant over that of the substrate, but this is not the case.

\begin{table*}
	\caption{\label{tab:dilution}Values of the dilution factors $D_\text{bulk}$ and dilution factor densities $\varepsilon_\text{bulk}$, $\varepsilon_\text{shear}$, and $\varepsilon = \varepsilon_\text{bulk} + \varepsilon_\text{shear}$, for the low vibrational modes of a substrate with Poisson's ratio $\nu=0.16$. The modes are indexed by $(m,n)$, where $m$ (resp. $n$) stand for the numbers of radial (resp. angular, or azimuthal) nodes. Four cases are considered for the Poisson's ratio of the coating: $\nu=0.31$ (\ce{Ta2O5}), $\nu=0.30$ (\ce{TiO2}:\ce{Nb2O5}), $\nu=0.19$ (\ce{SiO2}), and $\nu=0.16$ (Corning \ce{SiO2}).}
	\begin{ruledtabular}
		\begin{tabular}{c|cccc|cccc|cccc|cccc}
			& \multicolumn{4}{c|}{$\nu=0.31$} & \multicolumn{4}{c|}{$\nu=0.30$}  & \multicolumn{4}{c|}{$\nu=0.19$} & \multicolumn{4}{c}{$\nu=0.16$} \\
			\colrule
			\addlinespace[1ex]
			Mode	&	$D_\text{bulk}$	&	$ \varepsilon $	&	$\varepsilon_\text{bulk}$	&	$\varepsilon_\text{shear}$	&	$D_\text{bulk}$	&	$\varepsilon$	&	$\varepsilon_\text{bulk}$	&	$\varepsilon_\text{shear}$	&	$D_\text{bulk}$	&	$\varepsilon$	&	$\varepsilon_\text{bulk}$	&	$\varepsilon_\text{shear}$	&	$D_\text{bulk}$	&	$\varepsilon$	&	$\varepsilon_\text{bulk}$	&	$\varepsilon_\text{shear}$	\\
			(0,2)	&	0.062	&	1.004	&	0.145	&	0.859	&	0.063	&	1.007	&	0.149	&	0.858	&	0.070	&	1.047	&	0.179	&	0.869	&	0.070	&	1.061	&	0.184	&	0.878	\\
			(0,3)	&	0.096	&	1.731	&	0.259	&	1.472	&	0.098	&	1.738	&	0.266	&	1.472	&	0.110	&	1.839	&	0.325	&	1.513	&	0.111	&	1.871	&	0.336	&	1.535	\\
			(0,4)	&	0.118	&	2.327	&	0.356	&	1.971	&	0.120	&	2.340	&	0.367	&	1.973	&	0.136	&	2.501	&	0.454	&	2.047	&	0.138	&	2.552	&	0.471	&	2.081	\\
			(0,5)	&	0.133	&	2.848	&	0.443	&	2.404	&	0.136	&	2.866	&	0.457	&	2.409	&	0.155	&	3.086	&	0.571	&	2.515	&	0.157	&	3.155	&	0.594	&	2.561	\\
			(0,6)	&	0.144	&	3.318	&	0.523	&	2.795	&	0.147	&	3.341	&	0.540	&	2.801	&	0.169	&	3.618	&	0.680	&	2.938	&	0.172	&	3.703	&	0.708	&	2.996	\\
			(0,7)	&	0.152	&	3.753	&	0.599	&	3.154	&	0.156	&	3.780	&	0.618	&	3.162	&	0.181	&	4.111	&	0.782	&	3.329	&	0.184	&	4.213	&	0.815	&	3.398	\\
			(0,8)	&	0.159	&	4.160	&	0.669	&	3.491	&	0.163	&	4.192	&	0.691	&	3.501	&	0.190	&	4.575	&	0.879	&	3.696	&	0.193	&	4.692	&	0.917	&	3.775	\\
			(0,9)	&	0.165	&	4.546	&	0.737	&	3.809	&	0.169	&	4.582	&	0.761	&	3.820	&	0.197	&	5.014	&	0.971	&	4.043	&	0.201	&	5.146	&	1.014	&	4.132	\\
			(0,10)	&	0.170	&	4.914	&	0.802	&	4.112	&	0.174	&	4.954	&	0.828	&	4.125	&	0.204	&	5.434	&	1.060	&	4.374	&	0.208	&	5.580	&	1.108	&	4.473	\\
			\addlinespace[1.2ex]
			(1,0)	&	0.328	&	0.491	&	0.090	&	0.401	&	0.340	&	0.495	&	0.093	&	0.402	&	0.443	&	0.546	&	0.121	&	0.425	&	0.464	&	0.561	&	0.127	&	0.434	\\
			(1,1)	&	0.263	&	0.306	&	0.023	&	0.284	&	0.271	&	0.310	&	0.023	&	0.287	&	0.337	&	0.349	&	0.029	&	0.320	&	0.349	&	0.360	&	0.030	&	0.330	\\
			(1,2)	&	0.235	&	0.434	&	0.002	&	0.432	&	0.242	&	0.440	&	0.002	&	0.438	&	0.295	&	0.506	&	0.002	&	0.504	&	0.304	&	0.524	&	0.002	&	0.522	\\
			(1,3)	&	0.222	&	0.623	&	0.002	&	0.621	&	0.229	&	0.631	&	0.002	&	0.629	&	0.277	&	0.720	&	0.002	&	0.718	&	0.284	&	0.744	&	0.002	&	0.742	\\
			(1,4)	&	0.216	&	0.816	&	0.012	&	0.804	&	0.222	&	0.826	&	0.013	&	0.813	&	0.268	&	0.933	&	0.015	&	0.917	&	0.275	&	0.961	&	0.016	&	0.945	\\
			(1,5)	&	0.213	&	1.001	&	0.028	&	0.973	&	0.219	&	1.012	&	0.029	&	0.983	&	0.264	&	1.132	&	0.035	&	1.097	&	0.271	&	1.164	&	0.036	&	1.128	\\
			(1,6)	&	0.212	&	1.175	&	0.047	&	1.128	&	0.218	&	1.187	&	0.048	&	1.139	&	0.262	&	1.318	&	0.058	&	1.260	&	0.268	&	1.353	&	0.059	&	1.293	\\
			(1,7)	&	0.211	&	1.338	&	0.067	&	1.271	&	0.217	&	1.351	&	0.069	&	1.282	&	0.261	&	1.491	&	0.083	&	1.408	&	0.268	&	1.528	&	0.085	&	1.443	\\
			(1,8)	&	0.211	&	1.492	&	0.088	&	1.404	&	0.217	&	1.506	&	0.090	&	1.416	&	0.261	&	1.653	&	0.109	&	1.544	&	0.268	&	1.692	&	0.112	&	1.581	\\
			(1,9)	&	0.212	&	1.639	&	0.109	&	1.529	&	0.217	&	1.653	&	0.112	&	1.541	&	0.262	&	1.807	&	0.135	&	1.672	&	0.268	&	1.848	&	0.139	&	1.709	\\
			(1,10)	&	0.212	&	1.779	&	0.131	&	1.648	&	0.218	&	1.793	&	0.135	&	1.659	&	0.262	&	1.953	&	0.162	&	1.791	&	0.269	&	1.995	&	0.166	&	1.829	\\
			\addlinespace[1.2ex]
			(2,0)	&	0.260	&	0.107	&	0.020	&	0.088	&	0.268	&	0.108	&	0.020	&	0.088	&	0.333	&	0.113	&	0.025	&	0.088	&	0.344	&	0.115	&	0.026	&	0.089	\\
			(2,1)	&	0.250	&	0.127	&	0.009	&	0.117	&	0.257	&	0.128	&	0.010	&	0.118	&	0.317	&	0.143	&	0.012	&	0.131	&	0.328	&	0.147	&	0.012	&	0.135	\\
			(2,2)	&	0.242	&	0.228	&	0.003	&	0.226	&	0.249	&	0.231	&	0.003	&	0.229	&	0.306	&	0.266	&	0.003	&	0.263	&	0.315	&	0.276	&	0.003	&	0.273	\\
			(2,3)	&	0.236	&	0.357	&	0.000	&	0.357	&	0.243	&	0.362	&	0.000	&	0.362	&	0.297	&	0.418	&	0.000	&	0.418	&	0.306	&	0.433	&	0.000	&	0.433	\\
			(2,4)	&	0.232	&	0.492	&	0.001	&	0.491	&	0.239	&	0.498	&	0.001	&	0.497	&	0.292	&	0.573	&	0.001	&	0.572	&	0.300	&	0.593	&	0.001	&	0.592	\\
			(2,5)	&	0.230	&	0.624	&	0.005	&	0.619	&	0.236	&	0.632	&	0.005	&	0.627	&	0.288	&	0.723	&	0.006	&	0.717	&	0.296	&	0.748	&	0.006	&	0.742	\\
			(2,6)	&	0.228	&	0.750	&	0.010	&	0.740	&	0.234	&	0.760	&	0.011	&	0.749	&	0.285	&	0.865	&	0.013	&	0.852	&	0.293	&	0.894	&	0.013	&	0.880	\\
			(2,7)	&	0.226	&	0.870	&	0.018	&	0.853	&	0.233	&	0.881	&	0.018	&	0.863	&	0.283	&	0.999	&	0.022	&	0.977	&	0.291	&	1.031	&	0.023	&	1.008	\\
			(2,8)	&	0.226	&	0.984	&	0.026	&	0.958	&	0.232	&	0.996	&	0.027	&	0.969	&	0.282	&	1.124	&	0.033	&	1.092	&	0.290	&	1.159	&	0.034	&	1.125	\\
			(2,9)	&	0.225	&	1.092	&	0.036	&	1.056	&	0.231	&	1.105	&	0.037	&	1.068	&	0.281	&	1.242	&	0.044	&	1.198	&	0.289	&	1.279	&	0.046	&	1.234	\\
			(2,10)	&	0.224	&	1.194	&	0.046	&	1.149	&	0.231	&	1.208	&	0.047	&	1.161	&	0.280	&	1.354	&	0.057	&	1.297	&	0.288	&	1.393	&	0.059	&	1.334	\\
			\addlinespace[1.2ex]
			(3,0)	&	0.249	&	0.045	&	0.008	&	0.037	&	0.256	&	0.046	&	0.009	&	0.037	&	0.316	&	0.048	&	0.011	&	0.037	&	0.326	&	0.048	&	0.011	&	0.037	\\
			(3,1)	&	0.245	&	0.068	&	0.005	&	0.063	&	0.253	&	0.069	&	0.005	&	0.063	&	0.311	&	0.076	&	0.006	&	0.070	&	0.321	&	0.078	&	0.007	&	0.072	\\
			(3,2)	&	0.242	&	0.137	&	0.002	&	0.135	&	0.249	&	0.139	&	0.002	&	0.137	&	0.306	&	0.160	&	0.003	&	0.157	&	0.316	&	0.166	&	0.003	&	0.163	\\
			(3,3)	&	0.239	&	0.228	&	0.0004	&	0.227	&	0.247	&	0.231	&	0.0004	&	0.230	&	0.302	&	0.267	&	0.0005	&	0.266	&	0.311	&	0.276	&	0.001	&	0.276	\\
		\end{tabular}
	\end{ruledtabular}
\end{table*}
The bulk dilution factors $D_{\text{bulk}}=E_{\text{bulk}}/E$ values are reported in table \ref{tab:dilution} for some modes. In figure \ref{fig:Dbulkshear} the values of the bulk and shear dilution factor are plotted, considering $D_\text{shear}=1-D_\text{bulk}$. It can be observed that the modes having only angular nodes, therefore those whose deformation is mostly at the edge, are largely dominated by shear deformations.

\section{Dilution factors for the edge}
\label{app:Dedge}
 When considering the coating, there are three edge dilution factors that have to be calculated for each $w_{m,n}$:
\begin{equation}
	\varepsilon=\lim_{\ell\to 0}\frac{R}{\ell}D_\text{edge}=\lim_{\ell\to 0}\frac{R}{\ell}\frac{E_\text{edge}}{E_d}\medspace,
	\label{eq:varepsilon}
\end{equation}
and
\begin{equation}
	\varepsilon_\text{bulk}=\lim_{\ell\to 0}\frac{R}{\ell}\frac{E^\text{edge}_\text{bulk}}{E_d}\medspace,\qquad\varepsilon_\text{shear}=\lim_{\ell\to 0}\frac{R}{\ell}\frac{E^\text{edge}_\text{shear}}{E_d}\medspace.
	\label{eq:varepsilon_bulk}
\end{equation}
In each expression the denominator $E_d$ comes from the integration of the total infinitesimal energy \ref{eq:energydisk} over the whole disk ($0\leq\theta\leq2\pi$ ; $0\leq r\leq R$).
On the contrary, the numerator always comes from an integration performed all around the annular edge ($0\leq\theta\leq2\pi$ ; $r=R$) of thickness $\int dr=\ell$.
For $E_B$ we consider the total energy expression \ref{eq:energydisk}, whereas for $E^\text{edge}_\text{bulk}$ we consider the expression \ref{eq:Freeenergybulk_final} related to the dilatation elastic energy only. Using the expression for the shear elastic energy we have verified that the shear energy at the edge $E^\text{edge}_\text{shear}$ (not reported here) is exactly $E_\text{edge}-E^\text{edge}_\text{bulk}$. The values of the three edge dilution factors are listed in table \ref{tab:dilution} for some modes.
In figure \ref{fig:Dbulkshearedge} the dilution factor $\varepsilon$ at typical resonance frequency for a 75 mm silica disk 1 mm thick is shown. 
It can be observed from the figure that $\varepsilon$ follows a trend in which one can identify the separation of modes in families. Each family is identified by the radial node number m as shown in figure \ref{fig:ansys}. In each family the higher angular node number the more the vibration is confined to the edge.

The values of the bulk and shear dilution factor evaluated at the edge of a disk-shaped resonator are plotted in figure \ref{fig:Dbulkshearedge}. 


\bibliographystyle{elsarticle-num}
\bibliography{bibliography}

\end{document}